\documentclass{ws-mpla}

\usepackage[superscript]{cite}
\usepackage{amsmath,amssymb,color,epsfig}

\renewcommand{\draftnote}{}

\newcommand{\pb}{{\bf p}}

\renewcommand{\d}{{\mathrm{d}}}
\newcommand{\WW}{{\mathrm{WW}}}
\newcommand{\LIR}{{\mathrm{LIR}}}
\newcommand{\EOM}{{\mathrm{EOM}}}

\begin{document}

%
% Title, Authors, Affiliations 
% ============================
%
\markboth{T. Teckentrup, A. Metz, P. Schweitzer}
{Lorentz invariance relations and Wandzura-Wilczek approximation}

%%%%%%%%%%%%%%%%%%%%% Publisher's Area please ignore %%%%%%%%%%%%%%
\catchline{}{}{}{}{}
%%%%%%%%%%%%%%%%%%%%%%%%%%%%%%%%%%%%%%%%%%%%%%%%%%%%%%%%%%%%%%%%%%%

\title{LORENTZ INVARIANCE RELATIONS\\
AND WANDZURA-WILCZEK APPROXIMATION\footnote{
Contribution to the proceedings of the workshop ``Recent Advances in Perturbative QCD and Hadronic Physics'', July 20 -- 24, 2009, at ECT*, Trento (Italy).}}

\author{\footnotesize T. TECKENTRUP}

\address{Institut f{\"u}r Theoretische Physik II, Ruhr-Universit{\"a}t Bochum, 44780 Bochum, Germany\\
tobias.teckentrup@tp2.rub.de}

\author{A. METZ}

\address{Department of Physics, Barton Hall, Temple University, Philadelphia, PA 19122-6082, USA}

\author{P. SCHWEITZER}

\address{Department of Physics, University of Connecticut, Storrs, CT 06269, USA}

\maketitle

\pub{Received 14 October 2009}{~}

%
% Abstract
% ========
%
\begin{abstract}
A complete list of the so-called Lorentz invariance relations between parton distribution functions is given and some of their consequences are discussed, such as the Burkhardt-Cottingham sum rule.
The violation of these relations is considered in a model independent way. It is shown that several Lorentz invariance relations are not violated in a generalized Wandzura-Wilczek approximation, indicating that numerically their violation may be small. 

\end{abstract}

\ccode{PACS Nos.: 13.88.+e, 13.85.Ni, 13.60.-r}

%
% 1. Section: Introduction
% ========================
%
\section{Introduction}

Higher twist parton distribution functions (PDFs) and transverse parton momentum dependent ($\pb_T$-dependent) distribution functions (TMDs) contain important information on the partonic structure of the nucleon. They become more important because of the increasing accuracy of recent and planned high energy scattering experiments. The forward twist-3 PDFs are accessible through certain spin asymmetries in polarized inclusive deep inelastic lepton nucleon scattering (DIS) and Drell-Yan processes (integrated upon the transverse momentum of the dilepton pair)\cite{Jaffe:1991kp,Tangerman:1994bb,Adams:1994id,Abe:1998wq,Anthony:2002hy,Amarian:2003jy,Zheng:2004ce,Kramer:2005qe}. On the other hand the TMDs typically give rise to spin and azimuthal asymmetries in, for instance, semi-inclusive DIS\cite{Mulders:1995dh,Boer:1997nt,Barone:2001sp,Bacchetta:2006tn,D'Alesio:2007jt} and Drell-Yan\cite{Ralston:1979ys,Tangerman:1994eh,Boer:1999mm,Arnold:2008kf}, and significant effort has already been devoted to measure such observables\cite{Airapetian:1999tv,Avakian:2003pk,Airapetian:2004tw,Alexakhin:2005iw,Diefenthaler:2005gx,Ageev:2006da,Diefenthaler:2007rj,Alekseev:2008dn,Levorato:2008tv,Falciano:1986wk,Guanziroli:1987rp,Conway:1989fs,Zhu:2006gx}.

A systematic account of partonic $\pb_T$-effects leads to the 
introduction of many novel functions, making the interpretation 
of  especially twist-3 observables difficult. It was therefore appealing, when 
the so-called 'Lorentz invariance relations' (LIRs) were proposed\cite{Tangerman:1994bb,Mulders:1995dh,Boer:1997nt}. 
These relations were derived from
the Lorentz decomposition of the fully 
unintegrated correlator of two quark-fields, where the fields are located 
at arbitrary space-time positions.
However, in the original treatment 
\cite{Tangerman:1994bb,Mulders:1995dh,Boer:1997nt}
certain Lorentz-structures related solely to the
Wilson line in the correlator were not taken into consideration. 
After hints from model calculations\cite{Kundu:2001pk,Schlegel:2004rg} 
this was shown in a model-independent way\cite{Goeke:2003az}.

In this paper, in Sec.~\ref{sec:LIRs}, we first provide a complete 
list of LIRs and discuss some of their consequences.
Then, in Sec.~\ref{sec:LIRs_WW}, we review previous work\cite{Metz:2008ib}
and show that, although they are in general incorrect in QCD, several LIRs 
hold in an approximation which consists in systematically neglecting 
quark-gluon-quark correlations and current quark mass terms.
This is similar in spirit to the Wandzura-Wilczek-approximation.
We discuss the utility of such {\sl approximate relations} among TMDs.
Section~\ref{sec:summary} contains the summary and conclusions.

%
% 2. Section: LIRs
% ===========================
%
\section{Lorentz invariance relations}
\label{sec:LIRs}

In order to discuss the LIRs we have to study the fully unintegrated 
quark-quark correlation function $\Phi$ for a spin-$\frac{1}{2}$ hadron\cite{Goeke:2005hb}
(for spin-0 hadrons see Refs.~\refcite{Goeke:2003az,Bacchetta:2004zf}). 
The general decomposition of the correlator consists of
32 matrix structures multiplied by 12 scalar functions $A_i$ and 20 scalar functions $B_i$, the latter being associated solely with matrix structures induced by the Wilson line in the correlator\cite{Metz:2008ib}. The TMDs are defined through Dirac traces of the $\pb_T$-dependent correlator\cite{Mulders:1995dh,Bacchetta:2006tn,Metz:2008ib,Goeke:2005hb} integrated over $p^-$.
Consequently the TMDs can be expressed through $p^-$-integrals upon specific linear combinations
of the scalar functions $A_i$ and $B_i$\cite{Metz:2008ib}.

In total there are 32 quark TMDs up to twist-4 which exactly agrees with the number of independent amplitudes $A_i$ and $B_i$. If one neglects the structures $B_i$ induced by the Wilson line, the correlator $\Phi$ merely consists of 12 matrix structures, namely those related to the functions $A_i$. In that case the number of TMDs is larger than the number of the amplitudes $A_i$. This feature gives rise to LIRs. Some of them were discussed in literature\cite{Tangerman:1994bb,Mulders:1995dh,Boer:1997nt}, but so far no complete list of LIRs has been presented. How many independent LIRs do we obtain? There are 24 TMDs up to twist-3 and 12 amplitudes $A_i$. Thus, there must be $24 - 12 = 12$ LIRs in total. Let us distinguish time-reversal even (T-even) and T-odd LIRs.

There are six twist-2 and eight twist-3 T-even TMDs but nine T-even amplitudes. Thus there must be five T-even LIRs, namely
\begin{eqnarray} 
\label{eq:LIR1} g_T(x) \; &\stackrel{\LIR}{=}& \; g_1(x) + \frac{\d}{\d x} g^{(1)}_{1T}(x)\, ,
\end{eqnarray}
\begin{eqnarray}
\label{eq:LIR2} h_L(x) \; &\stackrel{\LIR}{=}& \; h_1(x) - \frac{\d}{\d x} h^{\perp(1)}_{1L}(x) \, , \\
\label{eq:LIR3} h_T(x) \; &\stackrel{\LIR}{=}& \; - \frac{\d}{\d x} h^{\perp(1)}_{1T}(x) \, , \\
\label{eq:LIR4} g_L^\perp(x) + \frac{\d}{\d x} g_T^{\perp(1)}(x) \; &\stackrel{\LIR}{=}& \; 0 \, ,\\ 
\label{eq:LIR5}
h_T(x,\pb^2_T)-h_T^\perp(x,\pb^2_T) \; &\stackrel{\LIR}{=}& \; h^{\perp}_{1L}(x,\pb^2_T) \, .
\end{eqnarray}

There are two twist-2 and eight twist-3 T-odd TMDs but three T-odd amplitudes, to recall $A_4$, $A_5$, $A_{12}$. Thus there must be seven T-odd LIRs, namely
\begin{eqnarray} 
\label{eq:LIR6} 
f_T(x) \; &\stackrel{\LIR}{=}& \; - \frac{\d}{\d x} f^{\perp(1)}_{1T}(x) \, , \\
\label{eq:LIR7} 
h(x) \; &\stackrel{\LIR}{=}& \; - \frac{\d}{\d x} h^{\perp(1)}_1(x) \, ,\\
\label{eq:LIR8}
e_L(x) + \frac{\d}{\d x} e^{(1)}_T(x) \; &\stackrel{\LIR}{=}& \; 0\, ,\\ 
\label{eq:LIR9}
f^{\perp}_L(x,\pb^2_T) \; &\stackrel{\LIR}{=}& \; - f^{\perp}_{1T}(x,\pb^2_T) \, , 
\phantom{\frac{\d}{\d x}}\\
\label{eq:LIR10}
e_T^\perp(x,\pb^2_T) \; &\stackrel{\LIR}{=}& \; 0 \, , \phantom{\frac{\d}{\d x}}\\
\label{eq:LIR11}
f_T^\perp(x,\pb^2_T) \; &\stackrel{\LIR}{=}& \; 0 \, , \phantom{\frac{\d}{\d x}}\\
\label{eq:LIR12}
g^\perp(x,\pb^2_T) \; &\stackrel{\LIR}{=}& \; 0 \, .\phantom{\frac{\d}{\d x}}
\end{eqnarray}
Notice that~(\ref{eq:LIR5}) and~(\ref{eq:LIR9})--(\ref{eq:LIR12}) hold unintegrated over $\pb_T$.\footnote{In case of the LIRs~(\ref{eq:LIR1})--(\ref{eq:LIR4}) and~(\ref{eq:LIR6})--(\ref{eq:LIR8}), where only $x$-dependent functions enter, corresponding relations involving higher $\pb_T$-moments of the same functions can be derived, provided that these moments exist.} The functions on the {\it lhs} of the LIRs are subleading, those on the {\it rhs} leading twist. The 'trivial' LIRs~(\ref{eq:LIR10})--(\ref{eq:LIR12}) arise because the respective TMDs are due to the $B_i$ amplitudes only. Finally, the (1)-$\pb_T$-moments of TMDs are defined as
\begin{equation} \label{eq:moments}
g^{(1)}_{1T}(x) \; = \; \int \d^2 \pb_T \, \frac{\pb^2_T}{2M^2} \, g_{1T}(x,\pb^2_T) \, , \; \text{etc.}
\end{equation}

Now let us discuss some consequences of the LIRs which follow if we 
integrate those relations over $x$ in the region of $-1\le x \le 1$. 
Here TMDs at negative $x$ are understood as $(\pm1)$TMDs of anti-quarks, 
depending on the $C$-parity of the operator \cite{Mulders:1995dh}. 
For example, with $x$ positive: $g_1^u(-x)=g_1^{\bar u}(x)$, 
$g_{1T}^{(1) u}(-x)=-g_{1T}^{(1)\bar u}(x)$, etc. Defining
$g_2(x)=g_T(x)-g_1(x)$, we obtain in this way from LIR~(\ref{eq:LIR1})
\begin{equation}
\label{eq:LIR-consequence-1} \int^1_{-1} \d x\;g_2(x) \; \stackrel{\LIR}{=} \; C \, ,
\end{equation}  
where the constant $C$ is defined, with an $\varepsilon>0$, assuming the limes exists, as
\begin{equation}
\label{eq:LIR-consequence-1a}
C \; \equiv \; \lim\limits_{\varepsilon\to 0}
\biggl[g_{1T}^{(1) q}(-\varepsilon)-g_{1T}^{(1)q}(\varepsilon)\biggr] \; \equiv \;
\lim\limits_{\varepsilon\to 0}
\biggl[-g_{1T}^{(1) q}(\varepsilon)-g_{1T}^{(1)\bar q}(\varepsilon)\biggr] \, .
\end{equation}
Similar sum rules hold for
$g^{\perp}_L(x)$, $h_T(x)$, $f_T(x)$, $h(x)$, $e_L(x)$, 
$h^{\perp}_T(x) + h^{\perp}_{1L}(x)$, and $h_L(x)-h_1(x)\equiv\frac12h_2(x)$.
For higher Mellin moments $n\ge 1$ one obtains, for example, 
\begin{equation}
\label{eq:LIR-consequence-2} 
\int^1_{-1} \d x\;x^n g_2(x)\;\stackrel{\LIR}{=}\; -\,n\int^1_{-1} \d x\;x^{n-1}g^{(1)}_{1T}(x)\, , \; \text{etc.}
\end{equation}
%Eq.~(\ref{eq:LIR9}) gives rise to relations also for higher $\pb_T$-moments 
%assuming they exist.

Interestingly, LIRs and the {\sl additional assumption} that 
$g_{1T}^{(1)}(x)$ is continuous at $x=0$ imply the validity
of the Burkhardt-Cottingham sum rule (BC)\cite{Burkhardt:1970ti}.
Under similar {\sl assumptions} one recovers also a sum rule\cite{Jaffe:1991kp} for $h_2(x)$ which is analog to BC.

However, there is a priori no reason why 
$\lim_{\varepsilon\to0}g_{1T}^{(1)     q}(\varepsilon)\stackrel{?}{=}
-\lim_{\varepsilon\to0}g_{1T}^{(1)\bar q}(\varepsilon)$
should hold, implying that BC could be violated.
At this point it is worthwhile remarking that we arrived at this
conclusion (always assuming that LIRs are valid) only by
carefully treating the integral in the vicinity of $x=0$. 
The total derivative $\frac{\d}{\d x}g_{1T}^{(1)}(x)$ gives 
no contribution at $x=\pm1$ where TMDs vanish\cite{Brodsky:2006hj}. 
But there are two more implicit boundaries, hidden by the compact notation, 
at $x\to\pm0$. This is of importance for the lowest Mellin moment 
(\ref{eq:LIR-consequence-1}) but not for higher moments
(\ref{eq:LIR-consequence-2}).

In fact, a discontinuity in $g_{1T}^{(1)}(x)$ at $x=0$ means that 
$g_2(x)$ has a singularity of the type $C\,\delta(x)$ at $x=0$
with the coefficient $C$ as defined in (\ref{eq:LIR-consequence-1a})
due to Eq.~(\ref{eq:LIR1}).
A $\delta(x)$-function in $g_2(x)$ corresponds to a real constant 
term in a spin flip Compton amplitude which persists to high energy\cite{Fox:1969cc,Broadhurst:1974er}. This phenomenon is known in Regge 
theory as ``a $J=0$ fixed pole with non-polynomial residue''\cite{Jaffe:1996zw}.
The possibility that BC is violated, and more generally, that PDFs may
have $\delta$-function type singularities at $x=0$ attract continuous 
interest in literature\cite{Anikin:2001ge,Burkardt:2001iy,Efremov:2002qh,Schweitzer:2003uy,Accardi:2009nv,D'Alesio:2009kv}.

We stress that these conclusions follow here from LIRs, i.e.\ in a 
{\sl not complete}  treatment of TMDs (due to the omission of $B_i$ amplitudes).
Interestingly, this framework though incomplete contains sufficiently rich 
features such as to allow for a violation of BC or the analog 
sum rule involving $h_2(x)$.

In QCD, which is a gauge theory where the Wilson line is mandatory, 
neither the LIRs nor their consequences are expected to hold. 
However, in relativistic quark models without gauge field degrees of freedom 
the T-even LIRs must hold. 
% In practice, they provide therefore valuable cross 
% checks for the model results.

%
% 3. Section: LIRs in a generalized WW approximation
% ==================================================
%
\section{Lorentz invariance relations in a generalized Wandzura-Wilczek approximation}
\label{sec:LIRs_WW}

Knowing that the LIRs are violated it is now natural to ask to what extent they are violated {\em numerically}. If one gets an indication that the violation of the LIRs should be small, these relations and their consequences can still serve as a useful tool --- at least for qualitative studies 
of the partonic nucleon structure. In what follows we consider certain LIRs and their violation in a model independent way using a special approximation. 

\subsection{T-even case}

For the discussion of the two T-even LIRs~(\ref{eq:LIR1}) and~(\ref{eq:LIR2}) we recall the following relations between $\pb_T$-integrated T-even PDFs~\cite{Jaffe:1991kp,Wandzura:1977qf}
\begin{equation}
\label{eq:coll_pdf}
g_T(x) \; = \; \int^1_x \frac{\d y}{y} g_1(y) + \tilde{g}'_T(x) \, , \quad h_L(x) \; = \; 2x \int^1_x \frac{\d y}{y^2} h_1(y) + \tilde{h}'_L(x) \, ,
\end{equation}
where $\tilde{g}'_T(x)$ and $\tilde{h}'_L(x)$ denote (purely interaction dependent) quark-gluon-quark correlations and terms proportional to current quark masses. An explicit representation of these terms can be found, e.g., in Ref.~\refcite{Belitsky:1997ay} and partly also in Ref.~\refcite{Accardi:2009nv}. The relations~(\ref{eq:coll_pdf}) isolate ``pure twist-3 terms'' in the PDFs $g_T(x)$ and $h_L(x)$.\footnote{Here the underlying ``working definition'' of twist~\cite{Jaffe:1996zw} (a PDF is of ``twist $t$'' if its contribution to the cross section is suppressed, in addition to kinematic factors, by $1/Q^{t-2}$ with $Q$ denoting the hard scale of the process) differs from the strict definition of twist (mass dimension of the operator minus its spin).}

The remarkable experimental observation is that $\tilde{g}'_T(x)$ is consistent with zero within the error bars\cite{Adams:1994id,Abe:1998wq,Anthony:2002hy,Amarian:2003jy,Zheng:2004ce,Kramer:2005qe} and to good accuracy one has
\begin{equation} 
\label{eq:WW1}
g_T(x) \; \stackrel{\WW}{\approx} \; \int^1_x \frac{\d y}{y} \; g_1(y) \, ,
\end{equation}
which is the Wandzura-Wilczek (WW) approximation. In Ref.~\refcite{Accardi:2009nv} on the basis of the present data for the DIS structure function $g_2$ 
it was found that the WW-relation~(\ref{eq:WW1}) works with an accuracy of the order $15-40\%$, 
though more data are needed to ultimately settle the situation. For our purpose this accuracy is quite satisfactory. Lattice QCD\cite{Gockeler:2000ja,Gockeler:2005vw} and the instanton model of the QCD vacuum\cite{Balla:1997hf} support the approximation~(\ref{eq:WW1}).
Interestingly, the latter predicts also $\tilde{h}'_L(x)$ to be small\cite{Dressler:1999hc} such that
\begin{equation}
\label{eq:WW2}
h_L(x) \; \stackrel{\WW}{\approx} \; 2x \int^1_x \frac{\d y}{y^2} \; h_1(y) \, .
\end{equation}
This approximation is not yet tested in experiment \cite{Koike:2008du}.
Further discussions of the WW approximation in related and other contexts can be found in the literature\cite{Ball:1996tb,Blumlein:1996vs,Blumlein:1998nv,Kivel:2000rb,Radyushkin:2000ap,Anikin:2001ge,Teryaev:1995um}.

Now it is possible to show that the LIRs~(\ref{eq:LIR1}) and~(\ref{eq:LIR2}) are not violated if one generalizes the WW approximation. For this purpose we consider the following exact relations\cite{Mulders:1995dh,Bacchetta:2006tn} originating from the QCD equations of motion (EOM)
\begin{eqnarray} 
\label{eq:EOM1}
g^{(1)}_{1T}(x) \; &\stackrel{\EOM}{=}& \; x\;(g_T(x)-\tilde{g}_T(x)) - \frac{m}{M}\;h_1(x) \, , \\
\label{eq:EOM2}
h^{\perp(1)}_{1L}(x) \; &\stackrel{\EOM}{=}& \;-\frac{x}{2} \;(h_L(x)-\tilde{h}_L(x)) + \frac{m}{2M}\;g_{1L}(x) \, ,
\end{eqnarray}
with the (1)-moments of the TMDs as defined in Eq.~(\ref{eq:moments}). The functions $\tilde{g}_T(x)$ and $\tilde{h}_L(x)$ denote twist-3 
quark-gluon-quark correlations. In lightcone gauge these objects, like $\tilde{g}'_T$ and $\tilde{h}'_L$, represent matrix elements of the type $\langle \, | \bar{\Psi} A_T \Psi | \,\rangle$. One therefore can assume that these functions are small as well, although the explicit form of $\tilde{g}_T \; (\tilde{h}_L)$ differs from the one of $\tilde{g}'_T \; (\tilde{h}'_L)$. In the following we denote as ``WW-type approximation'' the neglect of the tilde-functions (and quark mass terms) in the EOM-relations.\footnote{In Ref.~\refcite{Bacchetta:2006tn} for instance this approximation  (for brevity) was just called ``WW approximation'' because, like Eqs.~(\ref{eq:WW1},~\ref{eq:WW2}), it also corresponds to neglecting purely interaction dependent terms.} Below we will also address the phenomenological justification of the WW-type approximation.

In order to proceed we introduce the measures $\Delta_{g}(x)$ and $\Delta_{h}(x)$ for 
the violation of the respective LIRs (see for instance 
Refs.~\refcite{Belitsky:1997ay,Accardi:2009nv} for explicit forms of these terms) according to
\begin{eqnarray} 
\label{eq:LIR1_delta}
g_T(x) \; &=& \; g_1(x) + \frac{\d}{\d x} g^{(1)}_{1T}(x) + \Delta_{g}(x) \, , \\
\label{eq:LIR2_delta}
h_L(x) \; &=& \; h_1(x) - \frac{\d}{\d x} h^{\perp(1)}_{1L}(x) + \Delta_{h}(x) \, .
\end{eqnarray}
If one substitutes the (1)-moments $g^{(1)}_{1T}$ and $h^{\perp(1)}_{1L}$ from~(\ref{eq:EOM1},~\ref{eq:EOM2}) in Eqs.~(\ref{eq:LIR1_delta},~\ref{eq:LIR2_delta}), and uses both the WW approximation and the WW-type approximation one finds
\begin{eqnarray} 
\label{eq:delta1}
\Delta_{g}(x) \; &\stackrel{\mathrm{WW,WW-type}}{\approx}& \; -g_1(x) - x \; \frac{\d}{\d x} \int^1_x \frac{\d y}{y} \; g_1(y) \;\;\; = \; 0 \, , \\
\label{eq:delta2}
\Delta_{h}(x) \; &\stackrel{\mathrm{WW,WW-type}}{\approx}& \; -h_1(x) - x^2 \; \frac{\d}{\d x} \int^1_x \frac{\d y}{y^2} \; h_1(y) \; = \; 0 \, .
\end{eqnarray}
Eqs.~(\ref{eq:delta1},~\ref{eq:delta2}) show that the LIRs~(\ref{eq:LIR1}) and~(\ref{eq:LIR2}) are not violated if a generalized WW approximation is applied. This result is not entirely surprising keeping in mind that the violation of the LIRs is related to the amplitudes $B_i$ which are associated with the gauge link of the fully unintegrated correlator. Therefore, the $B_i$'s are necessarily related to quark-gluon-quark correlations 
which are neglected in the (generalized) WW approximation. This is also in line with the fact that in relativistic nucleon models without gluonic degrees of freedom the LIRs must be fulfilled\cite{Jakob:1997wg,Efremov:2009ze}.

We also point out that in the generalized WW approximation instead of having the three functions $g_1$, $g_T$, and $g_{1T}^{(1)}$, there is only one 
independent PDF. The same applies to the $h$-functions. In particular, one can immediately write\cite{Avakian:2007mv}
\begin{eqnarray} 
\label{eq:mom_g}
g_{1T}^{(1)}(x) \; &\stackrel{\mathrm{WW,WW-type}}{\approx}& \; x \int^1_x \frac{\d y}{y} g_1(y) \, , \\
\label{eq:mom_h}
h_{1L}^{\perp(1)}(x) \; &\stackrel{\mathrm{WW,WW-type}}{\approx}& \; - x^2 \int^1_x \frac{\d y}{y^2} h_1(y) \, .
\end{eqnarray}
Phenomenological work on the basis of those relations was, for instance, carried out in Refs.~\refcite{Kotzinian:1995cz,Kotzinian:2006dw,Avakian:2007mv}. In Ref.~\refcite{Avakian:2007mv}, Eq.~(\ref{eq:mom_h}) was used in order to  describe data for a certain longitudinal single spin asymmetry in semi-inclusive DIS. This investigation shows that the approximation~(\ref{eq:mom_h}) is not excluded, although more precise data would be helpful for having a stronger test.

\subsection{T-odd case}

In case of the T-odd LIRs~(\ref{eq:LIR6})--(\ref{eq:LIR12}) the situation is slightly different and in principle even simpler. Due to time-reversal invariance the $\pb_T$-integrated T-odd PDFs $f_T(x)$, $h(x)$, and $e_L(x)$ vanish~\cite{Goeke:2005hb,Bacchetta:2006tn}, 
\begin{eqnarray} 
\label{eq:T_odd1}
f_T(x) \; &=& \; \int \d^2 \pb_T \; f_T(x,\pb^2_T) \; = \; 0 \, , \\
\label{eq:T_odd2}
h(x) \; &=& \; \int \d^2 \pb_T \; h(x,\pb^2_T) \; = \; 0 \, , \\
\label{eq:T_odd3}
e_L(x) \; &=& \; \int \d^2 \pb_T \; e_L (x,\pb^2_T) \; = \; 0 \, ,
\end{eqnarray}
which implies, considering the LIRs~(\ref{eq:LIR6})--(\ref{eq:LIR8}), that
\begin{equation}
\label{eq:LIR_T_odd}
\frac{\d}{\d x} \; f^{\perp(1)}_{1T}(x) \; \stackrel{\LIR}{=} \; 0 \, , \quad \frac{\d}{\d x} \; h^{\perp(1)}_1(x) \; \stackrel{\LIR}{=} \; 0 \, , \quad \frac{\d}{\d x} \; e^{(1)}_T(x) \; \stackrel{\LIR}{=} \; 0 \, .
\end{equation}
This means that $f^{\perp(1)}_{1T}(x)$, $h^{\perp(1)}_1(x)$, and $e^{\perp(1)}_T(x)$ are constants. In fact, since these moments have to vanish for $x=1$, one can conclude that they should vanish for the entire $x$-range. So far we did not use any approximation and only assumed that the LIRs~(\ref{eq:LIR6})--(\ref{eq:LIR12}) are not violated. Now let us explore the EOM relations\cite{Bacchetta:2006tn} which, keeping in mind 
Eqs. (\ref{eq:T_odd1})--(\ref{eq:T_odd3}), imply\cite{Boer:2003cm,Ma:2003ut,Ji:2006ub}
\begin{eqnarray} 
& & f^{\perp(1)}_{1T}(x) \; \stackrel{\EOM}{=} \; x \; \tilde{f}_T(x) \, , \quad h^{\perp(1)}_1(x) \; \stackrel{\EOM}{=} \; \frac{x}{2} \; \tilde{h}(x) \, , \quad x \; e^{(1)}_T(x) \; \stackrel{\EOM}{=} \; x \; \tilde{e}^{(1)}_T(x) \, , \nonumber \\
& & x \; f^{\perp(1)}_L(x) \; \stackrel{\EOM}{=} \; x \; \tilde{f}^{\perp(1)}_L(x) \, , \quad x \; f^{\perp(1)}_T(x) \; \stackrel{\EOM}{=} \; x \; \tilde{f}^{\perp(1)}_T(x) + f^{\perp(1)}_{1T}(x) \, , \nonumber \\
& & x \; e^{\perp}_T(x,\pb^2_T) \; \stackrel{\EOM}{=} \; x \; \tilde{e}^{\perp}_T(x,\pb^2_T) + \frac{m}{M} \; f^{\perp}_{1T}(x,\pb^2_T) \, , \nonumber \\
\label{eq:EOM_T_odd}
& & x \; g^{\perp}(x,\pb^2_T) \; \stackrel{\EOM}{=} \; x \; \tilde{g}^{\perp}(x,\pb^2_T) + \frac{m}{M} \; h^{\perp}_1(x,\pb^2_T) \, .
\end{eqnarray}
In the WW-type approximation the tilde-functions and quark mass terms are set to zero. It then follows directly from~(\ref{eq:EOM_T_odd}) that $f^{\perp(1)}_{1T}(x)$,  $h^{\perp(1)}_1(x)$, $e^{(1)}_T(x)$, $f^{\perp(1)}_L(x)$, $f^{\perp(1)}_T(x)$, $e^{\perp}_T(x,\pb^2_T)$, and $g^{\perp}(x,\pb^2_T)$ are zero\cite{Bacchetta:2006tn}. 
This is consistent with the results following from the LIRs in~(\ref{eq:LIR_T_odd}) and from the LIRs (\ref{eq:LIR9})--(\ref{eq:LIR12}). 
So the T-odd LIRs~(\ref{eq:LIR6})--(\ref{eq:LIR12}), or ($x$-times) (1)-moments formed from them, are also not violated in the WW-type approximation.

Also for the T-odd functions we already have some phenomenological input on the status of the WW-type approximation. Since a nonzero asymmetry, typically attributed to the Sivers effect, was found in the HERMES experiment\cite{Airapetian:2004tw,Diefenthaler:2005gx,Diefenthaler:2007rj}, in the case of T-odd PDFs this approximation seems to be violated. On the other hand the observed effect is not very large (of the order of few percent), and one should not expect WW-type approximations to work to a much better accuracy than that. Moreover, the Sivers effect studied at COMPASS is compatible with zero both for a deuteron as well as a proton target\cite{Alexakhin:2005iw,Ageev:2006da,Alekseev:2008dn,Levorato:2008tv}. Therefore, the current experimental situation is not in conflict with a rather small $\tilde{f}_T$.

%
% 4. Section: Summary
% ===================
%
\section{Summary}
\label{sec:summary}

We have given a complete list of LIRs between parton distributions and have discussed some of their consequences.  
We have studied the LIRs, known to be violated in general, with the aim to understand how strong this violation might be. It was found that several LIRs are {\em satisfied} in a generalized WW approximation in which one systematically neglects certain quark-gluon-quark correlations as well as quark mass terms. 
If such terms were small which is sometimes assumed,
that would mean that LIRs could provide useful approximations for unknown PDFs and TMDs whenever applicable. Our approximation goes beyond the successful ``standard WW approximation'' quoted in Eqs.~(\ref{eq:WW1}) and~(\ref{eq:WW2}). In particular, we also neglected purely interaction dependent terms which show up in relations originating from the QCD equations of motion (see also Ref.~\refcite{Bacchetta:2006tn}). We argued that there exists experimental evidence for the validity of the generalized WW approximation. On the other hand more (precise) data and tests are needed before a final conclusion can be reached. Only forthcoming data analyses and experiments at COMPASS, HERMES, and Jefferson Lab can ultimately reveal to what extent the generalized WW approximation, the LIRs, and their consequences provide useful approximations. Eventually, it is likely that the quality of the approximation depends on the particular case (function) under consideration. 

%
% Acknowledgments
% ===============
%
\section*{Acknowledgments}

This paper is dedicated to our respected colleague and friend Anatoli V.~Efremov 
on the occasion of his 75th birthday. The authors thank the organizers of workshop on 
``Recent Advances in Perturbative QCD and Hadronic Physics'', ECT*, Trento (Italy) for
the opportunity to report our work at this memorable event.
The work is partially supported by the Verbundforschung ``Hadronen und Kerne'' of the BMBF. A.~M. acknowledges the support of the NSF under Grant No. PHY-0855501. P.~S. is supported by DOE contract No. DE-AC05-06OR23177, under which Jefferson Science Associates, LLC operates Jefferson Lab. T.~T. is supported by the Cusanuswerk.

%
% References
% ==========
%


\begin{thebibliography}{10}

%\cite{Jaffe:1991kp}
\bibitem{Jaffe:1991kp}
  R.~L.~Jaffe, X.~D.~Ji,
  %``Chiral odd parton distributions and polarized Drell-Yan,''
  Phys.\ Rev.\ Lett.\  {\bf 67} (1991) 552,
  Nucl.\ Phys.\  B {\bf 375} (1992) 527.
  %%CITATION = PRLTA,67,552;%%

%\cite{Tangerman:1994bb}
\bibitem{Tangerman:1994bb}
  R.~D.~Tangerman and P.~J.~Mulders,
  %``Polarized twist - three distributions g(T) and h(L) and the role of
  %intrinsic transverse momentum,''
  arXiv:hep-ph/9408305.
  %%CITATION = HEP-PH/9408305;%%

%\cite{Adams:1994id}
\bibitem{Adams:1994id}
  D.~Adams {\it et al.}  [Spin Muon Collaboration (SMC)],
  %``Spin asymmetry in muon proton deep inelastic scattering on a
  %transversely-polarized target,''
  Phys.\ Lett.\  B {\bf 336} (1994) 125.
  %%CITATION = PHLTA,B336,125;%%

%\cite{Abe:1998wq}
\bibitem{Abe:1998wq}
  K.~Abe {\it et al.}  [E143 collaboration],
  %``Measurements of the proton and deuteron spin structure functions g1  and
  %g2,''
  Phys.\ Rev.\  D {\bf 58} (1998) 112003.
  %%CITATION = PHRVA,D58,112003;%%

%\cite{Anthony:2002hy}
\bibitem{Anthony:2002hy}
  P.~L.~Anthony {\it et al.}  [E155 Collaboration],
  %``Precision measurement of the proton and deuteron spin structure  functions
  %g2 and asymmetries A(2),''
  Phys.\ Lett.\  B {\bf 553} (2003) 18.
  %%CITATION = PHLTA,B553,18;%%

%\cite{Amarian:2003jy}
\bibitem{Amarian:2003jy}
  M.~Amarian {\it et al.}  [JLab E94-010 Collaboration],
  %``Q**2 evolution of the neutron spin structure moments using a He-3
  %target,''
  Phys.\ Rev.\ Lett.\  {\bf 92} (2004) 022301.
  %%CITATION = PRLTA,92,022301;%%

%\cite{Zheng:2004ce}
\bibitem{Zheng:2004ce}
  X.~Zheng {\it et al.}  [JLab Hall A Collaboration],
  %``Precision measurement of the neutron spin asymmetries and  spin-dependent
  %structure functions in the valence quark region,''
  Phys.\ Rev.\  C {\bf 70} (2004) 065207.
  %%CITATION = PHRVA,C70,065207;%%

%\cite{Kramer:2005qe}
\bibitem{Kramer:2005qe}
  K.~Kramer {\it et al.},
  %``The Q**2-dependence of the neutron spin structure function g2(n) at low
  %Q**2,''
  Phys.\ Rev.\ Lett.\  {\bf 95} (2005) 142002.
  %%CITATION = PRLTA,95,142002;%%

%\cite{Mulders:1995dh}
\bibitem{Mulders:1995dh}
  P.~J.~Mulders and R.~D.~Tangerman,
  %``The complete tree-level result up to order 1/Q for polarized
  %deep-inelastic leptoproduction,''
  Nucl.\ Phys.\ B {\bf 461} (1996) 197\newline
  [Erratum-ibid.\ B {\bf 484} (1997) 538].
  %%CITATION = HEP-PH 9510301;%%

%\cite{Boer:1997nt}
\bibitem{Boer:1997nt}
  D.~Boer and P.~J.~Mulders,
  %``Time-reversal odd distribution functions in leptoproduction,''
  Phys.\ Rev.\ D {\bf 57} (1998) 5780.
  %%CITATION = HEP-PH 9711485;%%

%\cite{Barone:2001sp}
\bibitem{Barone:2001sp}
  V.~Barone, A.~Drago, and P.~G.~Ratcliffe,
  %``Transverse polarisation of quarks in hadrons,''
  Phys.\ Rept.\  {\bf 359} (2002) 1.
  %%CITATION = PRPLC,359,1;%%

%\cite{Bacchetta:2006tn}
\bibitem{Bacchetta:2006tn}
  A.~Bacchetta {\it et al.},
  %``Semi-inclusive deep inelastic scattering at small transverse momentum,''
  JHEP {\bf 0702} (2007) 093.
  %%CITATION = JHEPA,0702,093;%%

%\cite{D'Alesio:2007jt}
\bibitem{D'Alesio:2007jt}
  U.~D'Alesio and F.~Murgia,
  %``Azimuthal and Single Spin Asymmetries in Hard Scattering Processes,''
  Prog.\ Part.\ Nucl.\ Phys.\  {\bf 61} (2008) 394.
  %%CITATION = PPNPD,61,394;%%

%\cite{Ralston:1979ys}
\bibitem{Ralston:1979ys}
  J.~P.~Ralston and D.~E.~Soper,
  %``Production Of Dimuons From High-Energy Polarized Proton Proton
  %Collisions,''
  Nucl.\ Phys.\  B {\bf 152} (1979) 109.
  %%CITATION = NUPHA,B152,109;%%

%\cite{Tangerman:1994eh}
\bibitem{Tangerman:1994eh}
  R.~D.~Tangerman and P.~J.~Mulders,
  %``Intrinsic transverse momentum and the polarized Drell-Yan process,''
  Phys.\ Rev.\  D {\bf 51} (1995) 3357.
  %%CITATION = PHRVA,D51,3357;%%

%\cite{Boer:1999mm}
\bibitem{Boer:1999mm}
  D.~Boer,
  %``Investigating the origins of transverse spin asymmetries at RHIC,''
  Phys.\ Rev.\  D {\bf 60} (1999) 014012.
  %%CITATION = PHRVA,D60,014012;%%

%\cite{Arnold:2008kf}
\bibitem{Arnold:2008kf}
  S.~Arnold, A.~Metz, and M.~Schlegel,
  %``Dilepton production from polarized hadron hadron collisions,''
  Phys.\ Rev.\  D {\bf 79} (2009) 034005.
  %%CITATION = ARXIV:0809.2262;%%

\bibitem{Airapetian:1999tv}
  A.~Airapetian {\it et al.}  [HERMES Collaboration],
  %``Observation of a single-spin azimuthal asymmetry in semi-inclusive pion
  %electro-production,''
  Phys.\ Rev.\ Lett.\  {\bf 84} (2000) 4047;
  %%CITATION = PRLTA,84,4047;%%
%\bibitem{Airapetian:2001eg}
  %A.~Airapetian {\it et al.}  [HERMES Collaboration],
  %``Single-spin azimuthal asymmetries in electroproduction of neutral pions  in
  %semi-inclusive deep-inelastic scattering,''
  Phys.\ Rev.\  D {\bf 64} (2001) 097101;
  %%CITATION = PHRVA,D64,097101;%%
%\bibitem{Airapetian:2002mf}
  %A.~Airapetian {\it et al.}  [HERMES Collaboration],
  %``Measurement of single-spin azimuthal asymmetries in semi-inclusive
  %electroproduction of pions and kaons on a longitudinally polarised deuterium
  %target,''
  Phys.\ Lett.\  B {\bf 562} (2003) 182;
  %%CITATION = PHLTA,B562,182;%%
%\bibitem{Airapetian:2005jc}
  %A.~Airapetian {\it et al.}  [HERMES Collaboration],
  %``Subleading-twist effects in single-spin asymmetries in semi-inclusive
  %deep-inelastic scattering on a longitudinally polarized hydrogen  target,''
  Phys.\ Lett.\  B {\bf 622} (2005) 14.
  %%CITATION = PHLTA,B622,14;%%

\bibitem{Avakian:2003pk}
  H.~Avakian {\it et al.}  [CLAS Collaboration],
  %``Measurement of beam-spin asymmetries for deep inelastic pi+
  %electroproduction,''
  Phys.\ Rev.\  D {\bf 69} (2004) 112004;
  %%CITATION = PHRVA,D69,112004;%%
%\bibitem{Airapetian:2006rx}
  A.~Airapetian {\it et al.}  [HERMES Collaboration],
  %``Beam-spin asymmetries in the azimuthal distribution of pion
  %electroproduction,''
  Phys.\ Lett.\  B {\bf 648} (2007) 164.
  %%CITATION = PHLTA,B648,164;%%

\bibitem{Airapetian:2004tw}
  A.~Airapetian {\it et al.}  [HERMES Collaboration],
  %``Single-spin asymmetries in semi-inclusive deep-inelastic scattering on a
  %transversely polarized hydrogen target,''
  Phys.\ Rev.\ Lett.\  {\bf 94} (2005) 012002.
  %%CITATION = HEP-EX 0408013;%%

\bibitem{Alexakhin:2005iw}
  V.~Y.~Alexakhin {\it et al.}  [COMPASS Collaboration],
  %``First measurement of the transverse spin asymmetries of the deuteron in
  %semi-inclusive deep inelastic scattering,''
  Phys.\ Rev.\ Lett.\  {\bf 94} (2005) 202002.
  %%CITATION = PRLTA,94,202002;%%

\bibitem{Diefenthaler:2005gx}
  M.~Diefenthaler,
  %``Transversity measurements at HERMES,''
  AIP Conf.\ Proc.\  {\bf 792} (2005) 933.
  %%CITATION = APCPC,792,933;%%

%\cite{Ageev:2006da}
\bibitem{Ageev:2006da}
  E.~S.~Ageev {\it et al.}  [COMPASS Collaboration],
  %``A new measurement of the Collins and Sivers asymmetries on a  transversely
  %polarised deuteron target,''
  Nucl.\ Phys.\  B {\bf 765} (2007) 31.
  %%CITATION = NUPHA,B765,31;%%

\bibitem{Diefenthaler:2007rj}
  M.~Diefenthaler  [HERMES Collaboration],
  %``HERMES measurements of Collins and Sivers asymmetries from a transversely
  %polarised hydrogen target,''
  arXiv:0706.2242 [hep-ex].
  %%CITATION = ARXIV:0706.2242;%%

%\cite{:2008dn}
\bibitem{Alekseev:2008dn}
  M.~Alekseev {\it et al.}  [COMPASS Collaboration],
  %``Collins and Sivers asymmetries for pions and kaons in muon-deuteron DIS,''
  Phys.\ Lett.\  B {\bf 673} (2009) 127.
  %%CITATION = PHLTA,B673,127;%%

%\cite{Levorato:2008tv}
\bibitem{Levorato:2008tv}
  S.~Levorato  [COMPASS Collaboration],
  %``Single Spin Asymmetries on a transversely polarised proton target at
  %COMPASS,''
  arXiv:0808.0086 [hep-ex].
  %%CITATION = ARXIV:0808.0086;%%

%\cite{Falciano:1986wk}
\bibitem{Falciano:1986wk}
  S.~Falciano {\it et al.}  [NA10 Collaboration],
  %``Angular Distributions Of Muon Pairs Produced By 194-Gev/C Negative Pions,''
  Z.\ Phys.\  C {\bf 31} (1986) 513.
  %%CITATION = ZEPYA,C31,513;%%

%\cite{Guanziroli:1987rp}
\bibitem{Guanziroli:1987rp}
  M.~Guanziroli {\it et al.}  [NA10 Collaboration],
  %``ANGULAR DISTRIBUTIONS OF MUON PAIRS PRODUCED BY NEGATIVE PIONS ON DEUTERIUM
  %AND TUNGSTEN,''
  Z.\ Phys.\  C {\bf 37} (1988) 545.
  %%CITATION = ZEPYA,C37,545;%%

%\cite{Conway:1989fs}
\bibitem{Conway:1989fs}
  J.~S.~Conway {\it et al.},
  %``Experimental Study Of Muon Pairs Produced By 252-Gev Pions On Tungsten,''
  Phys.\ Rev.\  D {\bf 39} (1989) 92.
  %%CITATION = PHRVA,D39,92;%%

%\cite{Zhu:2006gx}
\bibitem{Zhu:2006gx}
  L.~Y.~Zhu {\it et al.}  [FNAL-E866/NuSea Coll.],
  %``Measurement of angular distributions of Drell-Yan dimuons in p + d
  %interaction at 800-GeV/c,''
  Phys.\ Rev.\ Lett.\  {\bf 99} (2007) 082301.
  %%CITATION = PRLTA,99,082301;%%
  
%\cite{Kundu:2001pk}
\bibitem{Kundu:2001pk}
  R.~Kundu and A.~Metz,
  %``Higher twist and transverse momentum dependent parton distributions: A
  %light-front Hamiltonian approach,''
  Phys.\ Rev.\  D {\bf 65} (2002) 014009.
  %%CITATION = PHRVA,D65,014009;%%

%\cite{Schlegel:2004rg}
\bibitem{Schlegel:2004rg}
  M.~Schlegel and A.~Metz,
  %``On the validity of Lorentz invariance relations between parton
  %distributions,''
  arXiv:hep-ph/0406289.
  %%CITATION = HEP-PH/0406289;%%

%\cite{Goeke:2003az}
\bibitem{Goeke:2003az}
  K.~Goeke, A.~Metz, P.~V.~Pobylitsa, M.~V.~Polyakov,
  %``Lorentz invariance relations among parton distributions revisited,''
  Phys.\ Lett.\  B {\bf 567} (2003) 27.
  %%CITATION = PHLTA,B567,27;%%
  
%\cite{Metz:2008ib}
\bibitem{Metz:2008ib}
  A.~Metz, P.~Schweitzer, and T.~Teckentrup,
  %``Lorentz invariance relations between parton distributions and the
  %Wandzura-Wilczek approximation,''
  Phys.\ Lett.\  B {\bf 680} (2009) 141.
  %%CITATION = PHLTA,B680,141;%%
  
%\cite{Goeke:2005hb}
\bibitem{Goeke:2005hb}
  K.~Goeke, A.~Metz, and M.~Schlegel,
  %``Parameterization of the quark-quark correlator of a spin-1/2 hadron,''
  Phys.\ Lett.\ B {\bf 618} (2005) 90.
  %%CITATION = HEP-PH 0504130;%%
  
%\cite{Bacchetta:2004zf}
\bibitem{Bacchetta:2004zf}
  A.~Bacchetta, P.~J.~Mulders, and F.~Pijlman,
  %``New observables in longitudinal single-spin asymmetries in  semi-inclusive
  %DIS,''
  Phys.\ Lett.\  B {\bf 595} (2004) 309.
  %%CITATION = PHLTA,B595,309;%%

\bibitem{Burkhardt:1970ti}
  H.~Burkhardt and W.~N.~Cottingham,
  %``Sum Rules For Forward Virtual Compton Scattering,''
  Annals Phys.\  {\bf 56} (1970) 453.
  %%CITATION = APNYA,56,453;%%
  
%\cite{Brodsky:2006hj}
\bibitem{Brodsky:2006hj}
  S.~J.~Brodsky and F.~Yuan,
  %``Single transverse-spin asymmetries at large-x,''
  Phys.\ Rev.\  D {\bf 74} (2006) 094018.
  %%CITATION = PHRVA,D74,094018;%%

%\cite{Fox:1969cc}
\bibitem{Fox:1969cc}
  G.~C.~Fox and D.~Z.~Freedman,
  %``Compton scattering sum rules and their saturation,''
  Phys.\ Rev.\  {\bf 182} (1969) 1628.
  %%CITATION = PHRVA,182,1628;%%

%\cite{Broadhurst:1974er}
\bibitem{Broadhurst:1974er}
  D.~J.~Broadhurst, J.~F.~Gunion, and R.~L.~Jaffe,
  %``Deep inelastic scattering and static properties of the baryons in the quark
  %gluon model,''
  Phys.\ Rev.\  D {\bf 8} (1973) 566;\\
  %%CITATION = PHRVA,D8,566;%%
%\cite{Broadhurst:1973fr}
%	\bibitem{Broadhurst:1973fr}
% D.~J.~Broadhurst, J.~F.~Gunion and R.~L.~Jaffe,
  %``On The Finiteness Of Scaling Sum Rules,''
  Annals Phys.\  {\bf 81} (1973) 88.
  %%CITATION = APNYA,81,88;%%

%\cite{Jaffe:1996zw}
\bibitem{Jaffe:1996zw}
  R.~L.~Jaffe,
  %``Spin, twist and hadron structure in deep inelastic processes,''
  arXiv:hep-ph/9602236.
  %%CITATION = HEP-PH 9602236;%%

%\cite{Burkardt:2001iy}
\bibitem{Burkardt:2001iy}
  M.~Burkardt and Y.~Koike,
  %``Violation of sum rules for twist-3 parton distributions in QCD,''
  Nucl.\ Phys.\  B {\bf 632} (2002) 311.
  %%CITATION = NUPHA,B632,311;%%

%\cite{Efremov:2002qh}
\bibitem{Efremov:2002qh}
  A.~V.~Efremov and P.~Schweitzer,
  %``The chirally-odd twist-3 distribution (e**a)(x),''
  JHEP {\bf 0308} (2003) 006.
  %%CITATION = JHEPA,0308,006;%%

%\cite{Schweitzer:2003uy}
\bibitem{Schweitzer:2003uy}
  P.~Schweitzer,
  %``The chirally-odd twist-3 distribution function e(x)**a in the chiral
  %quark-soliton model,''
  Phys.\ Rev.\  D {\bf 67} (2003) 114010.
  %%CITATION = PHRVA,D67,114010;%%

%\cite{D'Alesio:2009kv}
\bibitem{D'Alesio:2009kv}
  U.~D'Alesio, E.~Leader, and F.~Murgia,
  %``On the importance of Lorentz structure in the parton model: target mass
  %corrections, transverse momentum dependence, positivity bounds,''
  arXiv:0909.5650 [hep-ph].
  %%CITATION = ARXIV:0909.5650;%%

\bibitem{Anikin:2001ge}
  I.~V.~Anikin and O.~V.~Teryaev,
  %``Wandzura-Wilczek approximation from generalized rotational nvariance,''
  Phys.\ Lett.\  B {\bf 509} (2001) 95.
  %%CITATION = PHLTA,B509,95;%%

%\cite{Accardi:2009nv}
\bibitem{Accardi:2009nv}
  A.~Accardi, A.~Bacchetta, and M.~Schlegel,
  %``What can we learn from the breaking of the Wandzura-Wilczek relation?,''
  AIP Conf.\ Proc.\  {\bf 1155} (2009) 35;\\
  %%CITATION = APCPC,1155,35;%%
%\cite{Accardi:2009au}
%\bibitem{Accardi:2009au}
  A.~Accardi, A.~Bacchetta, W.~Melnitchouk, and M.~Schlegel,
  %``What can break the Wandzura--Wilczek relation?,''
  arXiv:0907.2942 [hep-ph].
  %%CITATION = ARXIV:0907.2942;%%

  
%\cite{Wandzura:1977qf}
\bibitem{Wandzura:1977qf}
  S.~Wandzura and F.~Wilczek,
  %``Sum Rules For Spin Dependent Electroproduction: Test Of Relativistic
  %Constituent Quarks,''
  Phys.\ Lett.\  B {\bf 72} (1977) 195.
  %%CITATION = PHLTA,B72,195;%%

%\cite{Belitsky:1997ay}
\bibitem{Belitsky:1997ay}
  A.~V.~Belitsky,
  %``Leading-order analysis of the twist-3 space- and time-like cut  vertices in
  %QCD,''
  arXiv:hep-ph/9703432.
  %%CITATION = HEP-PH/9703432;%%

%\cite{Gockeler:2000ja}
\bibitem{Gockeler:2000ja}
  M.~Gockeler {\it et al.},
  %``A lattice calculation of the nucleon's spin-dependent structure function
  %g2 revisited,''
  Phys.\ Rev.\  D {\bf 63} (2001) 074506.
  %%CITATION = PHRVA,D63,074506;%%

%\cite{Gockeler:2005vw}
\bibitem{Gockeler:2005vw}
  M.~Gockeler {\it et al.},
  %``Investigation of the second moment of the nucleon's g1 and g2 structure
  %functions in two-flavor lattice QCD,''
  Phys.\ Rev.\  D {\bf 72} (2005) 054507.
  %%CITATION = PHRVA,D72,054507;%%

%\cite{Balla:1997hf}
\bibitem{Balla:1997hf}
  J.~Balla, M.~V.~Polyakov and C.~Weiss,
  %``Nucleon matrix elements of higher-twist operators from the instanton
  %vacuum,''
  Nucl.\ Phys.\  B {\bf 510} (1998) 327.
  %%CITATION = NUPHA,B510,327;%%
  
%\cite{Dressler:1999hc}
\bibitem{Dressler:1999hc}
  B.~Dressler and M.~V.~Polyakov,
  %``On the twist-3 contribution to h(L) in the instanton vacuum,''
  Phys.\ Rev.\  D {\bf 61} (2000) 097501.
  %%CITATION = PHRVA,D61,097501;%%
  
%\cite{Koike:2008du}
\bibitem{Koike:2008du}
  Y.~Koike, K.~Tanaka, and S.~Yoshida,
  %``Drell-Yan double-spin asymmetry A_{LT} in polarized p\bar{p} collisions:
  %Wandzura-Wilczek contribution,''
  Phys.\ Lett.\  B {\bf 668} (2008) 286.
  %%CITATION = PHLTA,B668,286;%%

%\cite{Ball:1996tb}
\bibitem{Ball:1996tb}
  P.~Ball and V.~M.~Braun,
  %``The $\rho$ Meson Light-Cone Distribution Amplitudes of Leading Twist
  %Revisited,''
  Phys.\ Rev.\  D {\bf 54} (1996) 2182.
  %%CITATION = PHRVA,D54,2182;%%

%\cite{Blumlein:1996vs}
\bibitem{Blumlein:1996vs}
  J.~Bl\"umlein and N.~Kochelev,
  %``On the twist 2 and twist 3 contributions to the spin-dependent
  % electroweak structure functions,''
  Nucl.\ Phys.\  B {\bf 498} (1997) 285.
  %%CITATION = NUPHA,B498,285;%%

%\cite{Blumlein:1998nv}
\bibitem{Blumlein:1998nv}
  J.~Bl\"umlein and A.~Tkabladze,
  %``Target mass corrections for polarized structure functions and new
  % sum rules,''
  Nucl.\ Phys.\  B {\bf 553} (1999) 427.
  %%CITATION = NUPHA,B553,427;%%

%\cite{Kivel:2000rb}
\bibitem{Kivel:2000rb}
  N.~Kivel, M.~V.~Polyakov, A.~Sch\"afer, and O.~V.~Teryaev,
  %``On the Wandzura-Wilczek approximation for the twist-3 DVCS amplitude,''
  Phys.\ Lett.\  B {\bf 497} (2001)~73.
  %%CITATION = PHLTA,B497,73;%%

%\cite{Radyushkin:2000ap}
\bibitem{Radyushkin:2000ap}
  A.~V.~Radyushkin and C.~Weiss,
  %``DVCS amplitude at tree level: Transversality, twist-3, and
  %factorization,''
  Phys.\ Rev.\  D {\bf 63} (2001) 114012.
  %%CITATION = PHRVA,D63,114012;%%

\bibitem{Teryaev:1995um}
  O.~V.~Teryaev,
  %``Twist - three in proton nucleon single spin asymmetries,''
  arXiv:hep-ph/0102296.
  %%CITATION = HEP-PH/0102296;%%
  
%\cite{Jakob:1997wg}
\bibitem{Jakob:1997wg}
  R.~Jakob, P.~J.~Mulders, and J.~Rodrigues,
  %``Modelling quark distribution and fragmentation functions,''
  Nucl.\ Phys.\  A {\bf 626} (1997) 937.
  %%CITATION = NUPHA,A626,937;%%

%\cite{Efremov:2009ze}
\bibitem{Efremov:2009ze}
  A.~V.~Efremov, P.~Schweitzer, O.~V.~Teryaev, and P.~Zavada,
  %``Transverse momentum dependent distribution functions in a covariant parton
  %model approach with quark orbital motion,''
  Phys.\ Rev.\ D {\bf 80} (2009) 014021.
  %%CITATION = ARXIV:0903.3490;%%

%\cite{Kotzinian:1995cz}
\bibitem{Kotzinian:1995cz}
  A.~M.~Kotzinian and P.~J.~Mulders,
  %``Longitudinal quark polarization in transversely polarized nucleons,''
  Phys.\ Rev.\  D {\bf 54} (1996) 1229.
  %%CITATION = PHRVA,D54,1229;%%

%\cite{Kotzinian:2006dw}
\bibitem{Kotzinian:2006dw}
  A.~Kotzinian, B.~Parsamyan, and A.~Prokudin,
  %``Predictions for double spin asymmetry A(LT) in semi inclusive DIS,''
  Phys.\ Rev.\  D {\bf 73} (2006) 114017.
  %%CITATION = PHRVA,D73,114017;%%

%\cite{Avakian:2007mv}
\bibitem{Avakian:2007mv}
  H.~Avakian {\it et al.},
  %``Are there approximate relations among transverse momentum dependent
  %distribution functions?,''
  Phys.\ Rev.\  D {\bf 77} (2008) 014023.
  %%CITATION = PHRVA,D77,014023;%%
  
%\cite{Boer:2003cm}
\bibitem{Boer:2003cm}
  D.~Boer, P.~J.~Mulders, and F.~Pijlman,
  %``Universality of T-odd effects in single spin and azimuthal asymmetries,''
  Nucl.\ Phys.\  B {\bf 667} (2003) 201.
  %%CITATION = NUPHA,B667,201;%%

%\cite{Ma:2003ut}
\bibitem{Ma:2003ut}
  J.~P.~Ma and Q.~Wang,
  %``On unique predictions for single spin azimuthal asymmetry,''
  Eur.\ Phys.\ J.\  C {\bf 37} (2004) 293.
  %%CITATION = EPHJA,C37,293;%%

\bibitem{Ji:2006ub}
  X.~Ji, J.~W.~Qiu, W.~Vogelsang, and F.~Yuan,
  %``A unified picture for single transverse-spin asymmetries in hard
  %processes,''
  Phys.\ Rev.\ Lett.\  {\bf 97} (2006) 082002.
  %%CITATION = PRLTA,97,082002;%%

\end{thebibliography}
\end{document}